\documentclass[twocolumn]{aastex631}

\usepackage{orcidlink}
\usepackage{graphics,epsf}
\usepackage[utf8]{inputenc}
\usepackage{amsmath}                
\usepackage{amsfonts}               
\usepackage{amssymb}                
\usepackage{epsfig}                 
\usepackage{graphicx}
\usepackage{float}
\usepackage{color}
\usepackage{multirow}               
\usepackage{hyperref}
\usepackage{xspace}

\hypersetup{
    colorlinks=true,
    linkcolor=red,   
    urlcolor=cyan}

\usepackage[para,online,flushleft]{threeparttable}

\usepackage[colorinlistoftodos]{todonotes}

\newcommand{\km}{{~\rm km}}
\newcommand{\s}{{~\rm s}}

\newcommand{\g}{{~\rm g}}
\newcommand{\G}{{~\rm G}}
\newcommand{\K}{{~\rm K}}

\begin{document}

\title{Comments on: "Almost All Carbon/Oxygen White Dwarfs Can Support Double Detonations"}


\author{Noam Soker\,\orcidlink{0000-0003-0375-8987}} 
\affiliation{Department of Physics, Technion, Haifa, 3200003, Israel;  soker@physics.technion.ac.il}

\begin{abstract}
I critically review some claims in the paper ``Almost All Carbon/Oxygen White Dwarfs Can Support Double Detonations'' (arXiv:2405.19417). The claim of that paper that the community converges on a leading scenario of type Ia supernovae (SNe Ia), the double detonation scenario, is wrong, as hundreds of papers in recent years study five and more different SN Ia scenarios and their channels. Moreover, the finding by that paper that the double detonation scenario with the explosion of the secondary white dwarf (WD; the mass-donor WD) is common, i.e., the triple-detonation channel and the quadruple-detonation sub-channel, implies highly no-spherical explosions. The highly non-spherical explosions contradict the morphologies of many SN Ia remnants. I find that the results of that paper strengthen the claim that the double detonation scenario (with its channels) might account for a non-negligible fraction of peculiar SNe Ia but only for a very small fraction (or non at all) of normal SNe Ia.   
\end{abstract}


\section{Introduction} 
\label{sec:intro}

The community is far from a consensus on the leading scenarios for type Ia supernovae (SNe Ia), as is evident from reviews and classifications in the last decade (\citealt{Maozetal2014, MaedaTerada2016, Hoeflich2017, LivioMazzali2018, Soker2018Rev, Soker2019Rev, WangB2018,  Jhaetal2019NatAs, RuizLapuente2019, Ruiter2020, Aleoetal2023, Liuetal2023, Vinkoetal2023}). 
There are five, six, or even more different SN Ia scenarios (e.g., \citealt{Liuetal2023, Soker2024Rev}), with even more channels within some scenarios. The community does not converge on one-leading, two-leading, or even three-leading scenarios. Hundreds of papers in the last five years intensively discussed five scenarios and their channels for both normal and peculiar SNe Ia. \cite{Liuetal2023} discuss in detail the single degenerate (SD) scenario while \cite{Soker2024Rev} argue for the core degenerate (CD) scenario and the double degenerate scenario with the merger to explosion delay (MED) time (the DD-MED scenario) for normal SNe Ia, namely, the group of lonely white dwarf (WD) scenarios; other scenarios might account for a fraction of peculiar SNe Ia. Some researchers prefer the double detonation (DDet) scenario, where a WD accretes helium from a companion, and the ignition of the helium (the first detonation) detonates the CO interior of the WD (the second detonation).   

Despite this obvious state of SN Ia scenarios and channels, \cite{Shenetal2024} open the Abstract of their paper ``\textit{Almost All Carbon/Oxygen White Dwarfs Can Support Double Detonations}'' with the statement \textit{``Double detonations of sub-Chandrasekhar-mass white dwarfs (WDs) in unstably mass-transferring double WD binaries have become a leading contender for explaining most, if not all, Type Ia supernovae.''} They then open their Introduction with \textit{``The identity of Type Ia supernova (SN Ia) progenitors remains an unsolved mystery (see, e.g., \citealt{Liuetal2023} for a recent review), but recent work appears to be converging towards a solution. There is now mounting evidence that most, if not all, SNe Ia are the result of detonations of sub-Chandrasekhar-mass white dwarfs (WDs) in double WD systems.''} Concerning the highly diverse view of the community, the above opening sentences deceive the unfamiliar readers. It is fine to present one view and preference, but it becomes unethical to present one preference as the view that the diverse community converges on. More importantly, I find the new results that \cite{Shenetal2024} present to add problems to the DDet scenario as one of the scenarios for normal SNe Ia. 

\section{The challenges of the double detonation scenario} 
\label{sec:dDet}

There are two strong arguments against the DDet scenario (e.g., review by \citealt{Soker2019Rev}) : (1) The DDet scenario predicts a highly non-spherical explosion, contrary to the morphology of a large fraction of SNe Ia remnants (SNRs Ia) that are globally spherical or axially symmetric. Some SNRs Ia do possess a large departure from sphericity. (2) The surviving helium-donor companion is not observed in SNRs Ia, and the observed number of surviving hypervelocity WDs (the mass donor companions) is much below the number of normal SNe Ia \citep{Igoshevetal2023}. By arguing that most explosions in the DDet scenario are actually the triple detonation scenario (as coined and simulated by \citealt{Papishetal2015}), \cite{Shenetal2024} solve the second problem and explain the nondetection of surviving companions. In the triple detonation channel of the DDet scenario, helium detonation on the surface of one WD triggers a CO detonation in the same WD. The exploding WD ejecta triggers the explosion of the mass-donor WD companion, either the detonation of helium of a helium WD or helium shell detonation that triggers a CO detonation of the secondary WD in the quadruple detonation sub-channel of the triple channel (a term coined by \citealt{Tanikawaetal2019}). Namely, both WDs explode. (In the words of \citealt{Shenetal2024}: \textit{``Our work suggests that a majority of SNe Ia may arise from the double detonations of both WDs in a merging double WD binary.''}). The triple and quadruple detonation channels result in an explosion similar to the DD scenario without a MED time.  

As I now argue, solving the problem of the surviving companion with the triple detonation channel of the DDet scenario or with the quadruple detonation sub-channel, \cite{Shenetal2024} make the predicted asymmetrical explosion problem more severe. In the spirit of their paper, \cite{Shenetal2024} completely ignore the most severe problem of the DDet scenario, namely, that the predicted asymmetrical explosion contradicts observations of many SNRs Ia (see review by \citealt{Soker2019Rev}). 
Polarization measurements of SNe Ia show that many have global spherical symmetry, although there are non-spherical metal distributions (e.g., \citealt{Cikotaetal2019, Hoeflichetal2023}), which is also compatible with deflagration to the detonation of WDs near Chandrasekhar mass, as in the single-degenerate, double-degenerate with MED time, and the core degenerate.  

The two-dimensional hydrodynamical simulations by \cite{Papishetal2015} demonstrate the non-spherical ejecta of the DDet scenario, including the triple detonation channel. In the case of the DDet scenario, when the secondary (mass donor) WD does not explode, it casts an ‘ejecta shadow’ behind it. This leads to a supernova remnant that has a clear departure from spherical symmetry on a large scale. The SNR will appear spherical only if observed along or near the axis connecting the two WDs. Even this is not accurate, as the two WDs orbit each other, which will lead to deviation from a symmetry axis. \cite{Papishetal2015} conducted two-dimensional simulations and, therefore, could not follow the orbital motion of the two stars. In a simple, non-hydrodynamical, calculation, \cite{BraudoSoker2024} demonstrate that the orbital motion during the explosion adds some non-spherical structure to the inner ejecta in the DDet scenario. The effect of the ejecta shadow is seen in more detail and clearer in three-dimensional simulations (e.g., \citealt{Tanikawaetal2018, Tanikawaetal2019}) 

\begin{figure*}[]
\begin{center}
\includegraphics[trim=3.0cm 14.0cm 0.5cm 2.0cm,scale=1.1]{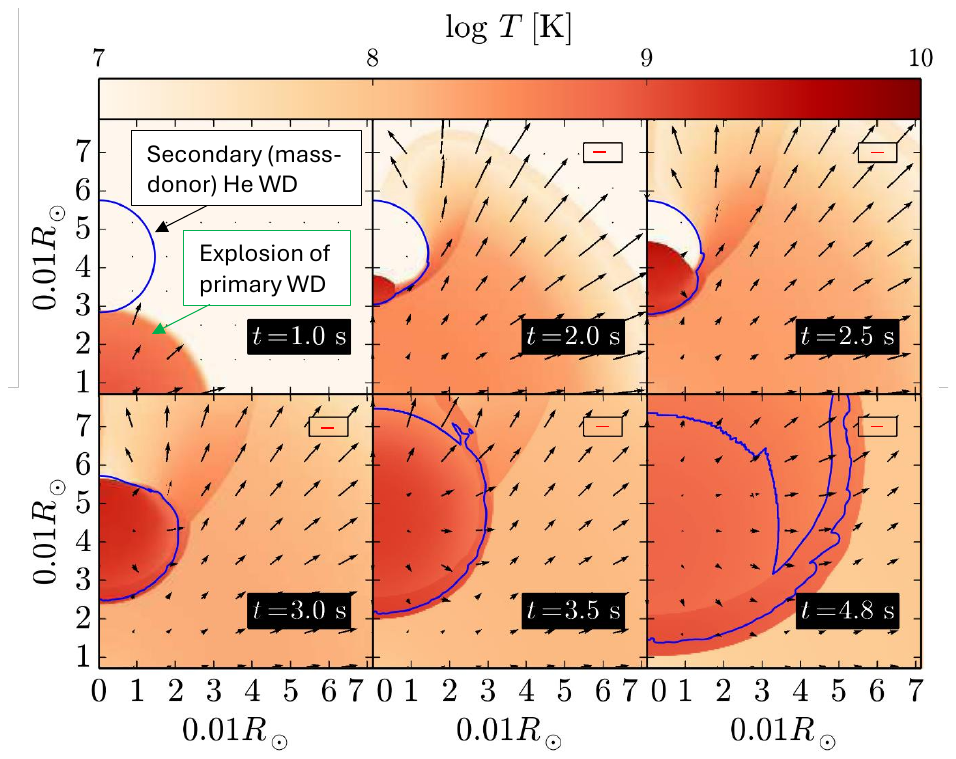} 
\caption{Temperature and velocity maps in the meridional plane at six times; figure built from a simulation by \cite{Papishetal2015} of the triple detonation channel of the DDet scenario. In this case, there is a He WD of $0.43 M_\odot$, the blue circle at the upper-left panel, at an initial distance of its center to the center of the explosion of $0.045 R_\odot$. 
Only one-quarter (not in full) of the meridional plane is shown. 
At $t = 2 \s$, the helium in the secondary WD is ignited (the high-temperature zone), and the secondary WD is exploded. The velocity at each grid point is proportional to the arrow length, with the inset showing a length corresponding to a velocity of $10,000 \km \s^{-1}$. Note the highly non-spherical explosion at the last time (bottom-left panel). }
\label{Fig:Figure1}
\end{center}
\end{figure*}

\cite{Papishetal2015} find that the triple-detonation channel of the DDet scenario leads to an even larger departure of the ejecta from global spherical symmetry than the DDet scenario. I demonstrate this in figure \ref{Fig:Figure1} of the temperature and velocity maps in the meridional plane of a simulation by \cite{Papishetal2015} of the triple-detonation channel. The simulation's high-temperature zone (red) is the material originating in the secondary WD. The temperature and velocity maps show a large-scale departure from spherical symmetry in the non-radial velocities and the unequal distribution of mass on the two sides (one of the primary and one of the secondary WD). \cite{Tanikawaetal2019}, in their three-dimensional simulations, follow the outflow in much more detail and to larger distances, one case of the triple-detonation channel and one of the quadruple-detonation sub-channel. Their density and composition maps present prominent large-scale departures from any symmetry, both in density and composition. 

The morphological expectations of large departure from axial symmetry and spherical symmetry of the triple-detonation channel and the quadruple-detonation sub-channel are in contradiction with the large fraction of SNRs Ia that possess global spherical structure, e.g., the Tycho SNR and SN 1006, and SNRs Ia that possess global axially-symmetric structures, e.g.,  SNR G1.9+0.3 (image by, e.g., \citealt{Enokiyaetal2023}). There are some SNRs Ia with large and global departures from any symmetry. I cannot rule out the DDet scenario for these SNRs Ia. It is upon the supporters of the DDet scenario (with the triple-detonation channel and quadruple-detonation sub-channel) to explain how their high non-spherical explosion can account for at least one of the observed SNR Ia. An example is the way the supporters of the core degenerate scenario account for the point-symmetric morphology of SNR G1.9+0.3 as a spherical explosion inside a planetary nebula (SNIP; \citealt{Soker2024G19}). 

\section{Summary} 
\label{sec:summary}

I conclude this comment on the recent paper by \cite{Shenetal2024} by reiterating my view that the DDet scenario and its triple-detonation channel and quadruple-detonation sub-channel can account at most for a small fraction of normal SNe Ia. 
The new results of \cite{Shenetal2024} reduce this fraction even more. Namely, I argue that the DDet scenario is not a dominant scenario of normal SNe Ia. It likely contributes to peculiar SNe Ia. My view is not in the consensus, as there is no consensus, and therefore, the claim by \cite{Shenetal2024}  that the community converges (in 2024) on a leading scenario is wrong. A consensus on the SN Ia leading scenario is for the future.    



\end{document}